\newcommand{\be}{\begin{equation}}\newcommand{\ee}{\end{equation}}
\newcommand{\bea}{\begin{eqnarray}}\newcommand{\eea}{\end{eqnarray}}
\newcommand{\nn}{\nonumber}\newcommand{\p}[1]{(\ref{#1})}
 \newcommand{\lb}[1]{\label{#1}}
 \newcommand\q{\quad}

\newcommand\s{\scriptscriptstyle}

\def\a{\alpha}

\def\bt{{\beta}}

\def\g{\gamma}

\def\de{\delta}

\def\eps{\epsilon}

\def\ve{\varepsilon}

\def\la{\lambda}

\def\om{\omega}

 \newcommand\te{\theta}  \def\bth{\bar\theta}

\def\rh{\rho}

\def\z{\zeta}

\def\J{\Psi}

\def\La{\Lambda}

\def\T{\Theta}

\def\pa{\partial}

\def\na{\nabla}

\newcommand\Tr{\mbox{Tr}\,}

\newcommand\ab{{\alpha\beta}}

\newcommand\A{{\s A}}

\newcommand\cD{{\cal D}}

\def\sfrac#1#2{{\textstyle\frac{#1}{#2}}}

\documentclass[12pt]{article}
\usepackage{amsfonts}
\def\theequation{\arabic{section}.\arabic{equation}}

\begin{document}

{\bf   CHERN-SIMONS THEORY IN THE $ SO(5)/U(2)$ HARMONIC SUPERSPACE}\\
\vspace{0.5cm}

{\bf B.M. Zupnik}\\

We consider the superspace of $D{=}3, N{=}5$ supersymmetry using
$SO(5)/U(2)$ harmonic coordinates. Three analytic $N{=}5$ gauge
superfields depend on three vector and six harmonic bosonic
coordinates and also on six Grassmann coordinates. Decomposition of
these superfields in Grassmann and harmonic coordinates yields
infinite-dimensional supermultiplets including a three-dimensional
gauge
Chern-Simons field and auxiliary bosonic and fermionic fields
carrying $SO(5)$ vector indices. The superfield action of this theory is
invariant with respect to $D{=}3, N{=}6$ conformal supersymmetry
realized on $N{=}5$ superfields.\\


\section{Introduction}

The three-dimensional gauge Chern-Simons (CS) theory \cite{ASch,JT}
is an
interesting example of the topological field theory closely
connected with  three-dimensional gravity, two-dimensional
conformal theories, and topological strings \cite{Wi1,Wi2}. In the
three-dimensional space with the coordinates $x^m $, the Chern-Simons
action has the form
  \bea
S_{CS}=\frac{k}{4\pi}\int d^3x\ve^{mnr}\Tr\{A_m(\pa_nA_r+ \sfrac{i}3
[A_n,A_r])\},\lb{SCS} \eea
where $m, n, r=0, 1, 2$, $\ve^{mnr}$ is the antisymmetric symbol,
and $A_m(x)$ is the non-Abelian gauge field. This action corresponds
to the classical solution with zero field-strength of the gauge field.
The functional $S_{CS}$ is not defined uniquely. It is invariant only
with respect to infinitesimal gauge transformations, but the quantization
of the constant $k$ guarantees the uniqueness of the quantity $\exp(iS_{CS})$
in the path integral of the quantum field theory.
We represent the Chern-Simons action as an integral
of the differential form
 \bea
S_{CS}=\frac{k}{4\pi}\int \Tr\{A d A+\sfrac{2i}3A^3\},\lb{SCS2}
\eea
where $A=dx^m A_m$ and $ d=dx^m\pa_m$. It follows from this representation
that the action is independent  on the metric of the three-dimensional space.
The term $S_{CS}$ can be regarded as an additional interaction of the gauge
field $A_m$ in the generalized three-dimensional Yang-Mills-Chern-Simons theory.
In this generalized
gauge theory, $ S_{CS}$ plays the role of the topological mass term.

Supersymmetric extensions of the $D{=}3$ Chern-Simons theory were discussed in
 \cite{Si}-\cite{Schw}. The simplest supersymmetric Chern-Simons theory was
constructed in the $D{=}3, N{=}1$ superspace with the real coordinates
$z=(x^m, \te^\a)$,
where the Grassmann coordinate $\te^\a$ has the spinor index $\a=1, 2$
of the group $SL(2,R)$ \cite{Si,Scho}. The superfield action of this theory is
defined on the gauge spinor superfield $A_\a(z)$, which contains the vector field
$A_m(x)$ and the spinor field $\psi_\a(x)$ in the adjoint representation of
the gauge group. In addition to \p{SCS}, the component action of the $N{=}1$
Chern-Simons theory includes the term $\int d^3x\Tr \psi^\a \psi_\a$. Hence,
the fermion $\psi_\a$ is an auxiliary field in the $N{=}1$ Chern-Simons theory,
although this field can describe physical degrees of freedom in the generalized
model with the additional supersymmetric Yang-Mills interaction.
The superfield action of the $D{=}3, N{=}1$ Chern-Simons theory can be interpreted
as the superspace integral of the differential Chern-Simons superform
$dA+\sfrac{2i}3A^3$ in the framework of our theory of superfield integral
forms \cite{ZP1}-\cite{Z6}.

The Abelian $D=3, N{=}2$ CS action was first constructed in the $D{=}3,
N{=}1$ superspace \cite{Si}. The corresponding non-Abelian action was considered
in the $D{=}3, N{=}2$ superspace with the help of the Hermitian superfield
$V(x^m,\te^\a,\bth^\a)$, where $\te^\a$ and $\bth^\a$ are the complex conjugated
spinor coordinates \cite{ZP1,Iv,NG}. In the  $N{=}2$
component-field Lagrangian,
bilinear terms with fermionic and scalar fields without
derivatives are added to the bosonic action. The unusual dualized form
 of the $N{=}2$ CS action contains the second vector field instead of the
scalar field\cite{NG}.

The $D{=}3, N{=}3$ CS theory was first analyzed by the
harmonic-superspace method \cite{ZK,Z3}. Using  the $SU(2)/U(1)$
harmonics \cite{GIKOS1} in the superspace with three spinor coordinates
$\te_\a^{ik}$ allows us to construct the analytic superspace with two
spinor coordinates $\te^{++}_\a$ and $ \te^0_\a$, where
the fourth supersymmetry can also be realized. The vector $N{=}3, 4$
supermultiplet is described
by the analytic superfield $V^{++}$ of  dimension zero, and the corresponding
analytic superfield strength $W^{++}=D^{++\a}D^{++}_\a V^{--}$ has the dimension
one and is expressed via spinor derivatives of the harmonic connection
$V^{--}$. It is convenient to build the superfield action of the $N{=}3$
Chern-Simons theory in the full $D{=}3, N{=}3$ by  analogy with the
action of the $D{=}4, N{=}2$ Yang-Mills theory in the corresponding
harmonic superspace \cite{Z2,GIOS}. The variation of the action of the
$CS_3^3$-theory can be presented in the full or analytic superspaces
\be
\de S_3^3=\int d^3x\,du\,d^6\te\Tr\{\de V^{++}V^{--}\}=
\int d^3x\,du\,d^{(-4)}\te\Tr\{\de V^{++}W^{++}\}=0,
\ee
where the measure  $d^{(-4)}\te$ contains derivatives in four analytic Grassmann
coordinates. We note that the equations of motion of the $CS_3^3$-theory
$W^{++}=0$ are covariant with respect to the fourth supersymmetry, although
the action possesses only three supersymmetries. The supersymmetric action of the
$D{=}3, N{=}4$ Yang-Mills theory can also be  constructed in the $D{=}3, N{=}3$
superspace, but the alternative formalism exists in the $N{=}4$ superspace
\cite{Z5}. The field-component form of the action of the $N{=}3$ Chern-Simons
theory was considered in \cite{KL,Kao}.

The component action of the $D{=}3, N{=}8$ Yang-Mills theory is obtained
by a dimensional reduction of the $D{=}4, N{=}4$ Yang-Mills theory. The action
16 supercharges is defined on the mass
shell of these theories.
A superfield version of the $D{=}4, N{=}3$ Yang-Mills theory $SYM_4^3$
was constructed in the $SU(3)/U(1)\times U(1)$ harmonic superspace
\cite{GIKOS2,GIOS};  this theory is invariant with respect to  12
supercharges only off the mass shell. It is easy to construct
a dimensional reduction of this Yang-Mills theory to the corresponding
three-dimensional superspace, but it is unclear whether the Chern-Simons
interaction can be included in this superspace. We should  say that the
integration measure
of the three-dimensional variant of the analytic $SU(3)/U(1)\times U(1)$
superspace has the dimension one.

 Using the $SO(5)/U(2)$ harmonic
superspace for the superfield description of the $D{=}3, N{=}5$ Chern-Simons
theory was proposed in \cite{HL}. An alternative formalism of this theory
 using the $SO(5)/U(1)\times U(1)$
harmonics and additional harmonic conditions was considered in \cite{Z8},
where it was
shown that the action of this model is invariant under the
$D{=}3, N{=}6$ superconformal group. It is useful to analyze
 the superfield formalism of the $N{=}6$ Chern-Simons theory in terms
of the $SO(5)/U(2)$ harmonics in detail.

In Sec. 2, we consider the $SO(5)/U(2)$ harmonics and the corresponding
harmonic derivatives. We define six Grassmann coordinates of the analytic superspace
using these harmonics and the ten Grassmann coordinates
of the full $N{=}5$ superspace. The coordinates of the analytic superspace
are considered real under a special conjugation. Three harmonic derivatives
preserve the Grassmann analyticity condition in the harmonic superspace.
The integration measure in the analytic $N{=}5$ superspace has the
dimension zero  compared with the  dimension two of the integration measure in the
full $D{=}3, N{=}5$ superspace. This property of the measure is very important
in  constructing  the superfield action of the theory.

In Sec. 3, we define the group of  analytic gauge transformations
using the harmonic condition of the $U(2)$ invariance for the superfield
parameters of these transformations. Basic gauge superfields of the theory
(prepotentials) satisfy the conditions of the Grassmann analyticity and
the harmonic $U(2)$  covariance. Reality conditions for the gauge group
and gauge superfields include the special conjugation of the analytic superspace.
We construct the superfield action of the theory  in the analytic
$D{=}3, N{=}5$ superspace \cite{HL,Z8}. One gauge superfield can be constructed
from two other prepotentials; the action of the theory contains two
independent superfields
in this representation. The classical equations of motion have only pure
gauge solutions for the prepotentials in analogy with the $N{=}1, 2, 3$
superfield Chern-Simons theories. The superfield action of the theory
is invariant under  transformations of the $D{=}3, N{=}5$
superconformal group realized on the prepotentials. Additional transformations
of the sixth supersymmetry are defined using the spinor derivative
preserving the Grassmann analyticity and $U(2)$  covariance.

In Sec. 4, we study the component structure of the $D{=}3, N{=}6$ Chern-Simons
theory.  We choose the supersymmetric gauge of the $U(2)$ covariant
prepotentials analogous to gauges of the Wess-Zumino (WZ) type  in other superfield
theories. The basic gauge superfield in this gauge includes the vector field
$A_m$ and the fermionic field $\psi_\a$ in the $SO(5)$ invariant sector
and also an infinite number of fermionic and bosonic fields with $SO(5)$
indices in the harmonic and Grassmann decompositions. The component Lagrangian
contains the Chern-Simons term for $A_m$ and also simple bilinear and trilinear
interactions of other bosonic and fermionic fields. Both the field strength of
the gauge field and all other fields vanish on the mass shell defined by
the classical equations of motion.

In Sec. 5, we discuss  another variant of the superfield action for the
Abelian $N{=}5$ prepotential with the coupling constant of a nontrivial
dimension. This action generates the ordinary Maxwell Lagrangian of the
vector field and other interactions of physical and auxiliary fields.
We consider the $D{=}3, N{=}5$ superconformal transformations in the full and analytic
superspaces  in   appendix. We also define  the superconformal
transformations of the integral measure and harmonic derivatives in the
analytic superspace.

The alternative formalism of the $D{=}3, N{=}6$ Chern-Simons theory
using the $SO(5)/U(1)\times U(1)$ harmonics was investigated in
 \cite{Z8}. This formalism assumes introducing  additional harmonic constraints
 for the parameters of the gauge group in accordance with
 the $U(2)$ covariance. The  $U(2)$ covariance is guaranteed in the
  $SO(5)/U(2)$ harmonic superspace. Another version of the $SO(5)/U(1)\times U(1)$
harmonic superspace without additional harmonic constraint was considered in
 \cite{Z7}.
 This model describes the
interaction of the  Chern-Simons supermultiplet with
unusual three-dimenasional matter fields.

\setcounter{equation}0
\section{$SO(5)/U(2)$ harmonic superspace}
The homogeneous space $SO(5)/U(2)$ is parametrized by elements of the harmonic
 5$\times$5 matrix
\be
U^K_a=(U^{+i}_a, U^0_a, U^-_{ia})
=(U^{+1}_a, U^{+2}_a, U^0_a, U^-_{1a}, U^-_{2a}),
\ee
where $a=1, \ldots 5$ is the vector index of the group $SO(5)$, $i=1, 2$ is
the spinor index of the group $SU(2)$, and $U(1)$-charges are denoted by symbols
$+, -, 0$. The basic relations for these harmonics are
\bea
&&U^{+i}_aU^{+k}_a=U^{+i}_aU^0_a=0,\q U^-_{ia}U^-_{ka}=U^-_{ia}
U^0_a=0,\q U^{+i}_aU^-_{ka}=
\de^i_k,\q U^0_aU^0_a=1,\nn\\
&&U^{+i}_aU^-_{ib}+U^-_{ia}U^{+i}_b+U^0_aU^0_b=\de_{ab}.
\eea

We consider the $SO(5)$ invariant harmonic derivatives with nonzero
$U(1)$ charges
\bea
&&\pa^{+i}=U^{+i}_a\frac{\pa}{\pa U^0_a}-U^0_a\frac{\pa}{\pa U^-_{ia}},\q
\pa^{+i}U^0_a=U^{+i}_a,\q \pa^{+i}U^-_{ka}=-\de^i_kU^0_a,\nn\\
&&\pa^{++}=U^+_{ia}\frac{\pa}{\pa U^-_{ia}},\q [\pa^{+i},\pa^{+k}]=
\ve^{ki}\pa^{++},\q\pa^{+i}\pa^{+}_i=\pa^{++},\nn\\
&&\pa_i^-=U^-_{ia}\frac{\pa}{\pa U^0_a}-U^0_a\frac{\pa}{\pa U^{+i}_a},
\q \pa_i^-U^0_a=U^-_{ia},\q \pa_i^-U^{+k}_a=-\de^k_iU^0_a,\\
&&\pa^{--}=U^{-i}_{a}\frac{\pa}{\pa U^{+i}_{a}},\q[\pa_i^-,\pa_k^-]=
\ve_{ki}\pa^{--},\q \pa^{-k}\pa_k^-=-\pa^{--},\nn
\eea
where some relations between these harmonic derivatives are defined.
The $U(1)$ neutral harmonic derivatives form the Lie algebra $U(2)$
\bea
&&\pa^i_k=U^{+i}_a\frac{\pa}{\pa U^{+k}_{a}}-U^{-}_{ka}\frac{\pa}
{\pa U^-_{ia}},\q[\pa^{+i},\pa^-_k]=-\pa^i_k,\\
&&\pa^0\equiv \pa^k_k=U^{+k}_a\frac{\pa}{\pa U^{+k}_{a}}-U^{-}_{ka}
\frac{\pa}{\pa U^-_{ka}},\q [\pa^{++},\pa^{--}]=\pa^0,\nn\\
&&\pa^i_k U^{+l}_a=\de_k^lU^{+i}_a,\q \pa^i_k U^-_{la}=-\de^i_lU^-_{ka}.
\eea
The operators $\pa^{+k}, \pa^{++}, \pa^-_k, \pa^{--}$ and $\pa^i_k$ satisfy
the commutation relations of the Lie algebra $SO(5)$.

An ordinary complex conjugation is defined on these harmonics
\bea
&&\overline{U^{+i}_{a}}=U^-_{ia},\q \overline{U^{0}_{a}}=U^{0}_{a},
\eea
but it is more convenient to use a special conjugation in the
harmonic space
\bea &&(U^{+i}_a)^\sim=U^{+i}_a,\q
(U^-_{ia})^\sim=U^-_{ia},\q (U^{0}_{a})^\sim=U^{0}_{a}.
\eea
All
harmonics are real with respect to this conjugation. On harmonic
polynomials with complex coefficients $f(U)$, the conjugation $\sim$
acts on these coefficients only, for example,
$(zU^K_a)^\sim=\bar{z}U^K_a$, where $\bar{z}$ is the complex
conjugation. The special conjugation acts on derivatives of the
harmonic functions and allows defining  the corresponding Hermitian
conjugation $\dagger$ of the harmonic operators
\bea
&&(\pa^{+i}f)^\sim=\pa^{+i}\tilde{f},\q
(\pa^{+i})^\dagger=-\pa^{+i},\q
(\pa^-_if)^\sim=\pa^-_i\tilde{f},\q (\pa^-_i)^\dagger=-\pa^-_i,\nn\\
&&(\pa^{\pm\pm}f)^\sim=\pa^{\pm\pm}\tilde{f},\q(\pa^{\pm\pm})^\dagger=
-\pa^{\pm\pm},
\eea
where $\pa^{\pm}_if=[\pa^{\pm}_i,f],\q \pa^{\pm\pm}f=[\pa^{\pm\pm},f]$.

The harmonic functions on the homogeneous 6-dimensional space
 $SO(5)/U(2)$ satisfy  the $U(2)$-invariance condition
\be
\pa^i_k f(U)=0.
\ee

The full superspace of the $D{=}3, N{=}5$ supersymmetry has the spinor
$CB$ coordinates $\te^\a_a,\q (\a=1, 2,\q a=1, 2, 3, 4, 5)$ in
addition to the coordinates $x^m$ of the three-dimensional Minkowski
space. The group $SL(2,R)\times SO(5)$ acts on the spinor coordinates.
We consider the superconformal transformations of these coordinates  in
appendix.

The $SO(5)/U(2)$ harmonics allow constructing projections of the spinor
coordinates and the partial spinor derivatives
\bea
&&\te^{+i\a}=U^{+i}_a\te^\a_a,\q \te^{0\a}=U^0_a\te^\a_a,\q
\te^{-\a}_i=U^-_{ia}\te^\a_a,\\
&&\pa^-_{i\a}=\pa/\pa\te^{+i\a},\q \pa^0_\a=\pa/\pa\te^{0\a},\q
\pa^{+i}_\a=\pa/\pa\te^{-\a}_i.\nn
\eea

The analytic coordinates ($AB$-representation) in the full harmonic superspace
use these projections of ten spinor coordinates
$\te^{+i\a}, \te^{0\a}, \te^{-\a}_i$ and the following representation
of the vector coordinate:
\bea
&&x^m_\A\equiv y^m=x^m+i(\te^{+k}\g^m\te^-_k)=x^m+i(\te_a\g^m\te_b)
 U^{+k}_aU^-_{kb}.
\eea

The analytic coordinates are real under the special conjugation
\bea
&&(\te^{+i\a})^\sim=\te^{+i\a},\q(\te^{0\a})^\sim=\te^{0\a},\q
(\te^{-\a}_i)^\sim=\te^{-\a}_i,\q(y^m)^\sim=y^m,\nn\\
&&(\Phi\Lambda)^\sim=\Lambda^\sim\Phi^\sim,\q (z\Lambda)^\sim=
\bar{z}\Lambda^\sim,\lb{spconj}
\eea
where we also define the action of the $\sim$-conjugation on arbitrary
superfields $\Phi, \Lambda$\q  ($z, \bar{z}$ are conjugate complex numbers).

We define the harmonic derivatives  in the $AB$ representation:
\bea
&&\cD^{+k}=\pa^{+k}-i(\te^{+k}\g^m\te^{0})\pa_m+\te^{+k\a}\pa^0_\a
-\te^{0\a}\pa^{+k}_\a,\nn\\
&&\cD^{++}=\pa^{++}+i(\te^{+k}\g^m\te^+_k)\pa_m+\te^{+\a}_k\pa^{+k}_{\a},
\nn\\
&&\cD^-_k=\pa^-_k+i(\te^-_k\g^m\te^0)\pa_m+\te^{-\a}_k\pa^0_\a-
\te^{0\a}\pa^-_{k\a},\lb{ABharm}\\
&&\cD^{--}=\pa^{--}-i(\te^{-}_k\g^m\te^{-k})\pa_m+\te^{-k\a}\pa^{-}_{k\a},
\nn\\
&&\cD^{k}_l=\pa^k_l+\te^{+k\a}\pa^-_{l\a}-\te^{-\a}_l\pa^{-k}_{\a}.\nn
\eea
These operators commute with  generators of the
 $N{=}5$ supersymmetry.

The $U(2)$ operator $\cD^{k}_l$ acts covariantly in the harmonic superspace
\bea
&&\cD^{k}_l\te^{+i\a}=\de^i_l\te^{+k\a},\q \cD^{k}_l\te^{-\a}_i=
-\de^k_i\te^{-\a}_l,\nn\\
&&[\cD^{k}_l,\cD^{+i}]=\de^i_l\cD^{+k},\q [\cD^{k}_l,\cD^{++}]=
\de^k_l\cD^{++},\q[\cD^{k}_l,\cD^-_{i}]=-\de^k_i\cD^-_{l}.
\eea

 The commutation relations are easily obtained:
\bea
&&[\cD^{+k},\cD^{+l}]=-\ve^{kl}\cD^{++},\q \cD^{+k}\cD^+_k=\cD^{++},\nn\\
&&[\cD^-_{k},\cD^-_{l}]=-\ve_{kl}\cD^{--},\q \cD^{-k}\cD^-_k=-\cD^{--},\nn\\
&&[\cD^-_l,\cD^{+k}]=\cD^k_l=\hat{\cD}^k_l+\sfrac12\de^k_l\cD^0,\q
[\cD^{++},\cD^{--}]=\cD^0,
\eea
where $\cD^0=\cD^k_k$ and $\hat{\cD}^k_l$ are the generators of
the subgroups $U(1)$ and $SU(2)$.

The relations between harmonic and spinor derivatives are
\bea
&&[\cD^{+i},D^0_\a]=D^{+i}_\a,\q [\cD^{+i},D^-_{k\a}]=-\de^i_kD^0_\a,\nn\\
&&[\cD^{++},D^0_\a]=0,\q [\cD^{++},D^-_{k\a}]=D^+_{k\a},\nn\\
&&[\cD^-_i,D^{+k}_\a]=-\de^i_kD^0_\a,\q [\cD^-_i,D^0_\a]=D^-_{i\a},\nn\\
&&[\cD^{--},D^{+k}_\a]=D^{-k}_\a,\q [\cD^{--},D^{0}_\a]=0,
\eea
where
\bea
&&D^{+i}_\a=\pa^{+i}_\a,\q D^-_{i\a}=\pa^-_{i\a}+2i\te^{-\bt}_i\pa_\ab,
\nn\\
&&D^0_\a=\pa^0_\a+i\te^{0\bt}\pa_\ab.
\eea

The coordinates of the analytic superspace $\z=(y^m, \te^{+i\a},
\te^{0\a},U_a^K)$ have the Grassmann dimension  6 and  dimension of the
even space  3+6. The functions  $\Phi(\z)$ satisfy the Grassmann analyticity
condition in this superspace
\be
D^{+k}_\a\Phi=0.
\ee
In addition to this condition, the analytic superfields in the
$SO(5)/U(2)$ harmonic superspace also have the $U(2)$-covariance.
For the  $U(2)$-scalar
superfields, this subsidiary condition is
\be
\cD^k_l\La(\z)=0.\lb{U2a}
\ee
For the $SU(2)$ spinor superfield with the $U(1)$ charge $q$, the
$U(2)$ covariance condition has the  form
\bea
\cD^0 \Phi^{(q)}_k=q\,\Phi^{(q)}_k,\q[\hat\cD^i_j,\Phi^{(q)}_{k}]=
-\de^i_k\Phi^{(q)}_j+\sfrac12\de^i_j\Phi^{(q)}_k.\lb{U2b}
\eea

The integration measure in the analytic superspace $d\mu^{(-4)}$ has
the dimension zero
\bea
&&d\mu^{(-4)}=dUd^3x_\A(\pa^0_\a)^2(\pa^-_{i\a})^4=dUd^3x_\A d\te^{(-4)},
\lb{meanal}
\eea
and is imaginary under the special conjugation
\be
(d\mu^{(-4)})^\sim=-d\mu^{(-4)}.
\ee
Integrals over the Grassmann and harmonic variables have the simple properties
\bea
&&\int d\te^{(-4)}(\te^0)^2\T^{(+4)}=1,\q\int dU=1,\q \int dU U^0_aU^0_b
=\sfrac15\de_{ab},\nn\\
&&\int dU U^{+k}_aU^-_{lb}=\sfrac15\de^k_l\de_{ab},
\eea
where combinations of the spinor coordinates $(\te^0)^2$ and $\T^{(+4)}$ are defined
in the appendix.

\setcounter{equation}0
\section{Gauge Chern-Simons theory in the
 $SO(5)/U(2)$ harmonic superspace}
 \subsection{Superfield action}

The harmonic derivatives  $\cD^{+k}$ and $\cD^{++}$ together with the spinor
derivatives $D^{+k}_\a$ determine the $CR$-structure of the harmonic
 $SO(5)/U(2)$ superspace. This $U(2)$-covariant $CR$-structure is
invariant under the $N{=}5$ supersymmetry. The $CR$-structure
should be preserved in the superfield gauge theory.

The gauge superfields (prepotentials) $V^{+k}(\z)$ and $V^{++}(\z)$ in the
harmonic $SO(5)/U(2)$ superspace satisfy  the
Grassmann analyticity and $U(2)$-covariance conditions
\bea
 D^{+k}_\a V^{+k}=D^{+k}_\a V^{++}=0,\q
\cD^i_jV^{+k}=\de^k_jV^{+i},\q \cD^i_jV^{++}=\de^i_jV^{++}.
\lb{U2V}
\eea
In the gauge group $SU(n)$, these traceless matrix superfields are anti-Hermitian
\be
(V^{+k})^\dagger=-V^{+k},\q (V^{++})^\dagger=-V^{++},
\ee
where operation $\dagger$ includes  transposition and the special conjugation.

The analytic superfield parameters of the gauge group $SU(n)$ satisfy
 the generalized $CR$ analyticity conditions
\be
D^{+k}_\a\La=\cD^i_j\La=0,\lb{CR}
\ee
and are traceless and anti-Hermitian $\La^\dagger=-\La$. We can use the
decomposition of these superfield parameters and prepotentials in terms of
generators of the adjoint representation of the gauge group $T_A$, for instance,
\be
V^{+k}=V^{+k}_AT_A,\q [T_A,T_B]=if_{ABC}T_C,\q \Tr(T_AT_B)=\de_{AB},
\ee
where $f_{ABC}$ are the group structure constants.

We treat these prepotentials as connections in the covariant gauge derivatives
\bea
&&\na^{+i}=\cD^{+i}+V^{+i},\q \na^{++}=\cD^{++}+V^{++},\nn\\
&&\de_\La V^{+i}=\cD^{+i}\La +[\La,V^{+i}],\q
\de_\La V^{++}=\cD^{++}\La +[\La,V^{++}],\lb{Vtrans}\\
&&D^{+k}_\a\de_\La V^{+k}=D^{+k}_\a\de_\La V^{++}=0,\q
\cD^i_j\de_\La V^{+k}=\de^k_j\de_\La V^{+i},\q \cD^i_j\de_\La V^{++}
=\de^i_j\de_\La V^{++},\nn
\eea
where the infinitesimal gauge transformations of the gauge superfields are
defined. These covariant derivatives commute with the spinor derivatives
$D^{+k}_\a$ and preserve the $CR$-structure in the harmonic superspace.

We can construct three analytic superfield strengths off the mass shell
\bea
&&F^{++}=\na^{++}+\sfrac12\ve_{ki}[\na^{+i},\na^{+k}]=V^{++}-\cD^{+k}V^{+}_k-
V^{+k}V^+_k,\nn\\
&&F^{(+3)k}=[\na^{++},\na^{+k}]=\cD^{++}V^{+k}-\cD^{+k}V^{++}+
[V^{++},V^{+k}],\lb{ancurv}\\
&&\de_\La F^{++}=[\La,F^{++}],\q \de_\La F^{(+3)k}=[\La,F^{(+3)k}].\nn
\eea

The superfield action in the analytic $SO(5)/U(2)$ superspace is defined
on three prepotentials $V^{+k}$ and $V^{++}$ by analogy with the off-shell
action of the $SYM_4^3$ theory \cite{GIKOS2}
\bea
S_1=\frac{ik}{12\pi}\int d\mu^{(-4)}\Tr\{V^{+j}\cD^{++}V^+_j
+2V^{++}\cD^+_jV^{+j}+(V^{++})^2+V^{++}[V^+_j,V^{+j}]\},\lb{act1}
\eea
where $k$ is the coupling constant and a numerical factor is chosen to
guarantee the correct normalization of the vector-field action. This action
is invariant under  the infinitesimal gauge transformations of
the prepotentials \p{Vtrans}. The idea of construction of the superfield
action in the harmonic $SO(5)/U(2)$ was proposed in \cite{HL}, although
the detailed construction of the superfield Chern-Simons theory was
not discussed there. An equivalent superfield action was considered
in the  alternative harmonic formalism \cite{Z8}.

The action $S_1$ yields superfield equations of motion meaning that
 the superfield strengths of the theory are trivial
\bea
&&F^{(+3)}_k=\cD^{++}V^+_k-\cD^+_kV^{++}+[V^{++},V^+_k]=0,\nn\\
&&F^{++}=V^{++}-\cD^{+k}V^{+}_k-V^{+k}V^+_k=0.\lb{equ}
\eea
These classical superfield equations have only pure gauge solutions for the
prepotentials
\be
V^{+k}=e^{-\La}\cD^{+k}e^\La,\q V^{++}=e^{-\La}\cD^{++}e^\La,
\ee
where $\La$ is an arbitrary analytic superfield. To prove this, we study
the corresponding classical field-component equations of motion in Sec. 4.
We note that formally similar superfield equations of the $SYM_4^3$ theory
have nontrivial solutions for the gauge superfields \cite{GIKOS2}. The main
reason for the differences in the character of the solutions of these two theories
is connected with the different Grassmann dimensions of the corresponding
analytic superspaces.

We rewrite action \p{act1} in the equivalent form
in terms of the prepotentials and analytic strengths  \p{ancurv}
\bea
S^\prime_1=\frac{ik}{12\pi}\int d\mu^{(-4)}\Tr\{V^{+k}F^{(+3)}_k+V^{++}
F^{++}+V^{++}V^{+k}V^+_k\}=\int d\mu^{(-4)}L^{+4},\lb{act1p}
\eea
where $L^{(+4)}$ is the superfield density of the action.

We consider the composite superfield
 \be
 \hat{V}^{++}=\cD^{+k}V^{+}_k+V^{+k}V^+_k,\q\de_\La\hat{V}^{++}=
 \cD^{++}\La +[\La,\hat{V}^{++}],
 \ee
 constructed from two other superfields. Substituting
$V^{++}\rightarrow \hat{V}^{++}$ in the action $S_1$,
 we can construct the alternative action on two independent
 prepotentials
\bea
&&S_2=\frac{ik}{12\pi}\int d\mu^{(-4)}\Tr\{V^{+j}\cD^{++}V^+_j
-(\hat{V}^{++})^2\}\nn\\
&&=\frac{ik}{12\pi}\int d\mu^{(-4)}\Tr\{V^{+j}\cD^{++}V^+_j
-(\cD^{+j}V^{+}_j+V^{+j}V^+_j)^2\}.\lb{act2}
\eea
This action corresponds to the equation of motion with the second-order
harmonic derivatives
\bea
&&\hat{F}^{(+3)}_k=\cD^{++}V^+_k-\cD^+_k\hat{V}^{++}
+[\hat{V}^{++},V^+_k]=0,
\eea
which is equivalent to equations \p{equ}.

\subsection{Superconformal invariance of the action}
The superconformal $SC_3^5$ transformations of the analytic coordinates
are considered in the appendix. Using  these transformations,
in particular, allows verifying  the superconformal invariance of the integration
measure $d\mu^{(-4)}$ in the analytic $SO(5)/U(2)$ superspace (see\p{scmu}).

We define the superconformal $SC_3^5$ transformations of the covariant
harmonic derivatives by analogy with formulas \p{scharm}
\be
\de_{sc}\na^{+k}=-\la^{+l}\cD^k_l,\q\de_{sc}\na^{++}
=\la^{++}\cD^k_k-\la^{+}_l\na^{+l},
\ee
where $\la_k^+, \la^{++}$ are  superfield parameters \p{la1}. The
corresponding transformations of the analytic prepotentials are
\bea
&&\de_{sc}V^{+k}=0,\q \de_{sc}V^{++}=-\la_k^+V^{+k}.
\eea

Analytic superfield strengths \p{ancurv} transform covariantly
 in the superconformal group
\bea
&&\de_{sc}F^{(+3)k}=\la^{+k}F^{++},\q \de_{sc}F^{++}=0.
\eea
Action \p{act1p} is invariant under the $SC_3^5$ transformations
\bea
\de_{sc}S^\prime_1=\frac{ik}{12\pi}\int d\mu^{(-4)}\Tr\{
V^{+k}\la^{+}_kF^{++}-\la_k^+V^{+k}F^{++}
-\la_l^+V^{+l}V^{+k}V^+_k\}=0,
\eea
where we use the identity
$$\Tr (V^{+l}V^{+k}V^+_k)=if_{ABC}V^{+l}_AV^{+k}_BV^+_{kC}=0. $$

The transformation of the sixth supersymmetry can be defined on the analytic
$N{=}5$ superfields
\bea
&&\de_6V^{++}=\eps^\a_6 D^0_\a V^{++},\q \de_6V^{+k}=\eps^\a_6 D^0_\a
V^{+k},\\
&&\de_6\cD^{+k}V^{+l}=\eps^\a_6 D^0_\a\cD^{+k}V^{+l},\q
\de_6\cD^{++}V^{+l}=\eps^\a_6 D^0_\a\cD^{++}V^{+l},\lb{6susy}
\eea
where $\eps^\a_6$ are the corresponding odd parameters. This transformation
preserves the Grassmann analyticity and $U(2)$-covariance
\be
\{D^0_\a,D^{+k}_\bt\}=0,\q [\cD^k_l,D^0_\a]=0,\q [\cD^{+k},D^0_\a]
=D^{+k}_\a,\q [\cD^{++},D^0_\a]=0.
\ee
The action \p{act1p} is invariant with respect to this sixth supersymmetry
\be
\de_6S^\prime_1=\int d\mu^{(-4)}\eps^\a_6 D^0_\a L^{(+4)}=0.
\ee
We can therefore speak on the $N{=}6$ Chern-Simons theory in the $N{=}5$ superspace.

\setcounter{equation}0
\section{Investigation of the component structure of the action}

The harmonic decomposition of the superfield gauge $U(1)$ parameter has the following
form:
\bea
&&\La=\La_0+\La_1+O(U^2),\q\cD^i_k\La=0,\nn\\
&&\La_0=ia+\te^{0\a}\bt_\a+(\te^0)^2d,\nn\\
&&\La_1=ia_bU^0_b+\te^{0\a}U^0_b\bt^b_\a+\te^{+k\a}U^-_{kb}\rh^b_\a
+(\te^0)^2U^0_ad^a+(\te^{+k}\te^0)U^-_{kb}f^b\nn\\
&&+(\te^{+k}\g^m\te^0)U^-_{kb}g^b_m+i(\te^0)^2\te^{+k\a}U^-_{kb}\pi^b_\a,
\eea
where  all local parameters  $a(y), a^b(y),
\bt_\a(y), \bt^b_\a(y)\ldots$ are chosen real to ensure the superfield condition
$\La^\sim=-\La$. In the group $SU(n)$, the corresponding local parameters are
Hermitian matrices with a zero trace. We also use conditions of the $U(2)$-invariance
$\cD^k_l\La_p=0$ in any degree $p$ of the harmonic decomposition. The local
parameters in $\La_0$ are  $SO(5)$ invariant and $\La_p$ contain parameters with
 $SO(5)$ tensor indices.

We  study the component decomposition of the action $S_2$ depending on two
prepotentials $V^{+k}$. In any gauge, these superfields can be decomposed
in degrees of harmonics
\bea
V^{+k}=V^{+k}_0+V^{+k}_1+O(U^2),
\eea
where each term contains Grassmann coordinates $\te^{0\a}, \te^{+k\a}$ and
the corresponding degrees of harmonics with the $U(2)$-covariance
condition $\cD^i_jV^{+k}=\de^k_jV^{+i}$ taken into account. The omitted terms
 contain an infinite
number of fields with tensor indices of the group $SO(5)$. We  note that
the $N{=}5$ supersymmetry transformations connect harmonic terms $V^{+k}_p$
with different degrees of harmonics. The supersymmetric $WZ$-gauge of each
harmonic term can be obtained using the transformations
$\de V^{+k}_p=\cD^{+k}\La_p$; for instance, the transformation
\be
\de V^{+k}_0=\cD^{+k}\La_0=\te^{+k\a}\bt_\a+(\te^{+k}\g^m\te^{0})
\pa_m a+\sfrac{i}2\te^{+k\g}(\te^{0})^2\pa_{\g\a}\bt^\a
+2(\te^{+k}\te^0)d
\ee
yields the gauge choice
\bea
V^{+k}_0=(\te^{+k}\g^m\te^{0})A_m+i(\te^0)^2\te^{+k\a}\psi_\a.\lb{WZ0}
\eea
The $WZ$-gauge of the term linear in harmonics  contains fields with the
$SO(5)$ vector index
\bea
&&V^{+k}_1=(\te^0)^2U^{+k}_aB_a+(\te^{+l}\g^m\te^{+}_l)
U^{-k}_aC_m^a\nn\\
&&+i\te^{+l(\a}\te^{+\bt}_l\te^{0\g)}U^{-k}_a\J^a_{(\a\bt\g)}
+i\te^{+l\a}\te^{+\bt}_l\te^{0}_\bt U^{-k}_a\xi^a_\a+i\te^{+k\bt}
\te^{+l}_\bt\te^{0\a} U^{-}_{la}\la^a_\a\nn\\
&&+i(\te^{+k}\te^{+l})(\te^0)^2U^-_{la}R_a
+i(\te^{+l}\g^m\te^{+}_l)(\te^0)^2U^{-k}_{a}G^a_m.\lb{O5vect}
\eea
We note that this decomposition contains the fermionic field
$\J^a_{(\a\bt\g)}$ with three spinor indices.

The harmonic-independent term of the superfield action \p{act2} is
\bea
&&S_0=\frac{ik}{12\pi}\int d\mu^{(-4)}\Tr\{V^{+j}_0\cD^{++}V^+_{0j}
-(\cD^{+j}V^{+}_{0j}+V^{+j}_0V^+_{0j})^2\},
\eea
where the gauge \p{WZ0} is chosen. The corresponding component Lagrangian
can be obtained by integrating over harmonics and spinor coordinates
\bea
L_0=\frac{k}{4\pi}\int d^3x\ve^{mnr}\Tr\{A_m(\pa_nA_r+\sfrac{i}3[A_n,A_r])+i
\psi^\a\psi_\a\}.
\eea
In addition to the standard Chern-Simons term, this Lagrangian contains
the bilinear interaction of the auxiliary fermion field.

Bilinear and trilinear interaction terms of auxiliary fields with an
arbitrary number of $SO(5)$ indices can also be  obtained from the superfield
action \p{act2} in the $WZ$-gauge, although these calculations are rather tedious.
On the mass shell, all auxiliary fields vanish and the gauge field  $A_m$ has
the  zero field-strength.

\setcounter{equation}0
\section{Alternative superfield action of  the $N{=}5$ theory}
In the three-dimensional $N=1,2$ and $3$ supersymmetries, the off-mass-shell
vector multiplets can be simultaneously used in the actions of the
supersymmetric Chern-Simons and Yang-Mills theories. A dimensional reduction
of the $D{=}4, N{=}3$ superfield gauge theory \cite{GIKOS2} to the
three-dimensional space yields the corresponding supersymmetric
$D{=}3, N{=}6$ Yang-Mills theory $SYM_3^6$ which preserves the automorphism group
$U(3)$ and breaks  the superconformal invariance. The equations of motion of
this theory have an additional supersymmetry \cite{Z9}. After an infinite number
of auxiliary fields are excluded, there remain the three-dimensional vector field,
seven scalar fields and eight two-component spinor fields, which form
the $D{=}3, N{=}8$
multiplet of the physical fields.  We do not know how to construct the
Chern-Simons action on the harmonic superfields of the $SYM_3^6$ theory.

We consider the Abelian gauge superfield in the
 $D{=}3, N{=}5$ supersymmetry
\be
\de_\La V^{+i}=\cD^{+i}\La.\lb{abel}
\ee
In addition to the Abelian version of the Chern-Simons action \p{act2},
we can also construct the alternative superfield action of the $U(1)$
gauge superfield
\bea
S_E=\frac{1}{12g^2}\int d\mu^{(-4)}(D^{0\a}D^0_\a V^{+k})(\cD^{++}V^+_k
-\cD^{+}_k\cD^{+l}V^+_l),\lb{SE}
\eea
where $g$ is a coupling constant of the dimension 1/2. This action is invariant
with respect to the gauge transformation  \p{abel}
\bea
&&\de_\La S_E~\sim~\int d\mu^{(-4)}(D^{0\a}D^0_\a \cD^{+k}\La)
(\cD^{++}V^+_k-\cD^{+}_k\cD^{+l}V^+_l)=0,
\eea
where we use the relations $D^{0\a}D^0_\a \cD^{+k}\La=\cD^{+k}
D^{0\a}D^0_\a \La, \q \cD^{++}=\cD^{+k}\cD^{+}_k$ and  integrate by parts.
The interaction $S_E$ breaks  the sixth supersymmetry
 \p{6susy} and also the conformal invariance.

In components, the action \p{SE} contains the Lagrangian of the three-dimensional
electrodynamics and also the kinetic term of the fermion field  $\psi^\a$.
In the sector of the $SO(5)$ vector fields \p{O5vect}, the action  $S_E$
yields bilinear terms of the bosonic and fermionic fields. Compared with
the supersymmetric  $N=1, 2, 4$ and 6 generalizations of electrodynamics,
the action \p{SE} contains an unusual interaction of the chirality 3/2
field $\J^{\a\bt\g}_a$.  In contrast to supergravity models, this field arises
 without  additional local supersymmetry here. The equation of motion for
the action $S_E$ has nontrivial solutions.

\setcounter{equation}0
\section{Conclusions}
We used the harmonic coordinates describing the $SO(5)/U(2)$ space in the
framework of the $D{=}3, N{=}5$ supersymmetry and investigated the
geometry of the gauge
theory in the analytic $SO(5)/U(2)$ superspace based on the Grassmann
analyticity and $U(2)$-covariance of the gauge superfields  in
detail. The gauge transformations of the superfield prepotentials of the theory
preserve the $CR$-structure of the spinor and harmonic covariant derivatives
in analogy with other examples of  gauge theories in harmonic superspaces.
We constructed  equivalent representations of the superfield action of the gauge theory
with three and two independent superfields. We showed that
these superfield actions are invariant under  the $N{=}5$
conformal supersymmetry and the additional sixth supersymmetry.
We studied the supersymmetric  $WZ$-gauge of the superfield prepotentials;
it defines an infinite number of the field components, including the vector
gauge field and auxiliary bosonic and fermionic fields. The superfield
equations of motion of this theory have only the pure gauge solutions.
The component equations of motion yield trivial (zero) solutions
for the field-strength of the vector field and all other auxiliary
fields in the Chern-Simons supermultiplet. The superfield formulation
can be useful in the study of the quantum properties of the
 $N{=}6$ Chern-Simons theory.

The author is grateful to E.A. Ivanov and P.S. Howe
for  interesting discussions. This work was partially supported by
the grants RFBR 06-02-16684, DFG 436 RUS 113/669-3, INTAS 05-10000008-7928
and by the Heisenberg-Landau programme.

\renewcommand\theequation{A.\arabic{equation}} \setcounter{equation}0
\section*{Appendix. Superconformal transformations}
{\bf  A.1. Superconformal  $CB$ transformations}\\

We use the real symmetric representation for the
$D{=}3$ gamma-matrices
\bea
&&(\g_m)^\ab=\ve^{\a\rh}(\g_m)^\bt_\rh=(\g_m)^{\bt\a},\q
(\g^m)_\ab(\g_m)^{\rh\g}=\de_\a^\rh\de_\bt^\g+\de_\bt^\rh\de_\a^\g,\nn\\
&&(\g_m\g_n)^\bt_\a=(\g_m)_\a^\rh(\g_n)_\rh^\bt=-(\g_m)_{\a\rh}
(\g_n)^{\rh\bt}=-\eta_{mn}\de^\bt_\a+\ve_{mnp}(\g^p)^\bt_\a,\\
&&(\g_m\g_n\g_r)^\bt_\a=-\ve_{mnr}\de^\bt_\a-\eta_{mn}(\g_r)^\bt_\a
-\eta_{nr}(\g_m)^\bt_\a+\eta_{mr}(\g_n)^\bt_\a,\nn
\eea
where $\eta_{mn}=\mbox{diag}(1,-1,-1)$ is the Minkowski metric, and
$\ve_{mnp}$ is the antisymmetric symbol in the $3D$ space. The spinor
$SL(2,R)$  indices are raised and lowed using the antisymmetric symbol
\bea
&&\psi^\a=\ve^{\ab}\psi_\bt,\q\psi_\a=\ve_{\ab}\psi^\bt,\q
\ve^{\ab}\ve_{\bt\g}=\de^\a_\g.
\eea
We use the  notation
\be
(\psi\xi)=\psi^\a\xi_\a,\q (\psi\g_m\xi)=\psi^\a(\g_m)_\ab\xi^\bt
\ee
for the scalar and vector combinations of
two spinors.

The bispinor representations of the three-dimensional coordinates and derivatives are
\bea
&&x^\ab=(\g_m)^\ab x^m,\q \pa_\ab=(\g^m)_\ab \pa_m
\eea
and the corresponding expression for the  $N{=}5$ spinor derivatives are
\bea
&&D_{a\a}=\pa_{a\a}+i\te^\bt_a\pa_\ab,\q\pa_{a\a}\te^\bt_b=\de_{ab}
\de^\bt_\a.
\eea

In the full superspace with the $CB$-coordinates $z=(x^m, \te^\a_a)$,
we define the  $N{=}5$ superconformal transformations
($SC_3^5$-transformations)
\bea
&&\de_{sc} x^\ab=c^\ab+\sfrac12a^\a_\g x^{\g\bt}+\sfrac12a^\bt_\g x^{\a\g}
+l\,x^\ab-i\eps^\a_b\te^\bt_b-i\eps^\bt_b\te^\a_b
+\sfrac12x^{\a\g}k_{\g\rh}x^{\rh\bt}-\sfrac12\te^\a_a\te^\g_a\te^\rh_b
\te^\bt_b k_{\g\rh}\nn\\
&&-\sfrac{i}2\eta_{a\g}\te^\a_ax^{\g\bt}-\sfrac{i}2\eta_{a\g}\te^\bt_a
x^{\g\a}+\sfrac12\eta_{a\rh}\te^\a_a\te^\rh_b\te^\bt_b
+\sfrac12\eta_{a\rh}\te^\bt_a\te^\rh_b\te^\a_b,\nn\\
&&\de_{sc}\te^\a_a=\sfrac12a^\a_\bt\te^\bt_a+\sfrac12l\,\te^\a_a
-\om_{ab}\te^\a_b+\sfrac12x^\ab\te^\g_a
k_{\bt\g}+\sfrac{i}2\te^\a_b\te^\bt_b\te^\g_a k_{\bt\g}+\eps^\a_a\nn\\
&&+\sfrac12x^\ab\eta_{a\bt}-i\te^\a_b\te^\g_a\eta_{b\g}+\sfrac{i}2\te^\a_b
\te^\bt_b\eta_{a\bt},
\eea
where $c^\ab=c^m(\g_m)^\ab,~ k^\ab=k^m(\g_m)^\ab,~a^\a_\bt= a^m
(\g_m)^\a_\bt$ and $l$ are the parameters of the $D{=}3$ conformal group,
$\om_{ab}$ are the $SO(5)$ parameters, and the Grassmann parameters
$\eps^\a_a$ and $\eta^\a_a$ correspond to the transformations of the
$Q$-supersymmetries and $S$-supersymmetries.

The generators of the $N{=}5$ superconformal group $SC_3^5$ in the $CB$-representation
are
\bea
&&P_m=\pa_m,\q Q^a_\a=\pa^a_\a-i\te^{a\bt}(\g^m)_\ab\pa_m,\q T_{ab}=
\te^\a_b\pa_{a\a}-\te^\a_a\pa_{b\a},\nn\\
&&L_m=\ve_{mnp}x^p\pa^n+\sfrac12\te^\a_a(\g_m)_\a^\bt\pa_\bt^a=
\sfrac12\ve_{mnp}L^{np},\q D=x^m\pa_m+\sfrac12\te^\a_a\pa_\a^a,\nn\\
&&K_m=x_mx^n\pa_n-\sfrac12x^2\pa_m-\sfrac14(\te_a\g_m\te_b)
(\te_a\g^n\te_b)\pa_n\nn\\
&&+\sfrac12x_m\te^\a_a\pa^a_\a-\sfrac12x^n\ve_{mnp}(\g^p)^\a_\g\te^\g_a
\pa^a_\a+\sfrac{i}2(\te_b\g_m\te_a)\te^\a_b\pa^a_\a,\\
&&S^a_\a=\sfrac{i}2\te_{a\a}x^m\pa_m-\sfrac{i}2\te^\bt_a\ve^{mnp}
(\g_p)_{\bt\a}x_n\pa_m+\sfrac12\te_{b\a}\te^\bt_a
(\g^m)_{\bt\rh}\te^\rh_b\pa_m\nn\\
&&-\sfrac12x^m(\g_m)_\a^\bt\pa_{a\bt}-i\te_{b\a}\te^\bt_a\pa_{a\bt}
+\sfrac{i}2\te_{b\a}\te^\bt_b\pa_{a\bt}.\nn
\eea
The local (active) transformations of superfields are determined by these
generators
\bea
\bar\de_{sc}\Phi=\de_{sc}\Phi-(c^mP_m+a^mL_m+\sfrac12\om_{ab}T_{ab}
+k^mK_m+lD+\eps^\a_aQ_{\a\a}+\eta^\a_aS_{\a\a})\Phi,
\eea
where $\de_{sc}\Phi$ are the coordinate (passive) superconformal transformations.

It is easy to obtain the (anti)commutator relations of the Lie
superalgebra $SC_3^5$
\bea
&&[L_m,L_n]=\ve_{mnp}L^p,\q[L_m,P_n]=\ve_{mnp}P^p,\q[L_m,K_n]=
\ve_{mnp}K^p,\q[L_m,D]=0,\nn\\
&&[L_m,Q_\a^a]=-\sfrac12(\g_m)^\bt_\a Q_\bt^a,\q[L_m,S_\a^a]=
-\sfrac12(\g_m)^\bt_\a S_\bt^a,\q[P_m,Q^a_\a]=0,\q[P_m,D]=P_m,\nn\\
&&[P_m,K_n]=\ve_{mnp}L^p+\eta_{mn}D,\q[P_m, S^a_\a]=
-\sfrac12(\g_m)^\bt_\a Q_\bt^a,\q[Q^a_\a,D]=\sfrac12Q^a_\a,\\
&&[K_m,Q_\a^a]=\sfrac12(\g_m)_\ab S^{a\bt},\q
\{Q_{a\a},S_{b\bt}\}=-\sfrac{i}2\de_{ab}(\g_m)_{\a\bt}L_m-i\de_{ab}
\ve_{\a\bt}D-i\ve_{\a\bt}T_{ab}\nn\\
&&\{Q^a_\a,Q^b_\bt\}=-2i\de^{ab}(\g^m)_\ab P_m,\q \{S^a_\a,S^b_\bt\}=
-i\de^{ab}(\g^m)_\ab K_m,\q [S^a_{\a},K_m]=0.\nn
\eea
\\

{\bf A.2. $SC^5_3$-transformations in analytic coordinates}\\

The superconformal $N{=}5$ transformations of the analytic coordinates
have the form
\bea
&&\de_{sc} x^\ab_\A=\de_{sc} y^\ab=c^\ab+L^\a_\g y^{\g\bt}+L^\bt_\g
y^{\a\g}+\sfrac12y^{\a\g}\,k_{\g\rh} y^{\bt\rh}+2l\,y^\ab
+i\om_{ab}U^{0}_aU^-_{kb}\te^{+k\a}\te^{0\bt}\nn\\
&&+i\om_{ab}U^{0}_aU^-_{kb}\te^{+k\bt}\te^{0\a}-i\eps^{0\a}\te^{0\bt}
-i\eps^{0\bt}\te^{0\a}-2i\eps^{-\a}_k\te^{+k\bt}-2i\eps^{-\bt}_k
\te^{+k\a}\nn\\
&&+\sfrac{i}2y^{\a\g}\te^{0\bt}\eta_\g^{0}+\sfrac{i}2y^{\bt\g}\te^{0\a}
\eta_\g^{0}+iy^{\a\g}\te^{+k\bt}\eta_{k\g}^{-}+iy^{\bt\g}\te^{+k\a}
\eta_{k\g}^{-},
\eea
\bea
&&\de_{sc}\te^{0\a}=\sfrac12a^\a_\bt\te^{0\bt}+l\,\te^{0\a}
+\sfrac12y^{\a\bt}\te^{0\g}k_{\bt\g}
-\om_{ab}U_a^{0}U^-_{kb}\te^{+k\a}  \nn\\&&
+\eps^{0\a}+\sfrac12y^{\a\bt}\eta^{0}_\bt
+\sfrac{i}4(\te^{0})^2\eta^{0\a}
-i\te^{+k\a}\te^{0\bt}\eta^{-}_{k\bt},
\eea
\bea
&&\de_{sc}\te^{+k\a}=\sfrac12a^\a_\bt\te^{+k\bt}+l\,\te^{+k\a}+
\sfrac12y^{\a\bt}\te^{+k\g}k_{\bt\g}
+\sfrac{i}4(\te^{0})^2\te^{+k\bt}k^\a_\bt \nn\\
&&-\om_{ab}U_a^{+k}U_{lb}^{-}
\te^{+l\a}
+\eps^{+k\a}+\sfrac12y^{\a\bt}\eta^{+k}_\bt
\nn\\
&&-i\te^{+l\a}\te^{+k\bt}\eta^{-}_{l\bt}+\sfrac{i}2\eta^{-\a}_l
(\te^{+k}\te^{+l})+\sfrac{i}4(\te^{0})^2\eta^{+k\a},
\eea
where we introduce the  notation
 $$
\eps^{0\a}=U^{0}_a\eps^\a_a,\q\eps^{\pm\a}_k=U^\pm_{ka}\eps^\a_a,\q
\eta^{0\a}=U^{0}_a\eta^\a_a,\q\eta^{\pm\a}_k=U^\pm_{ka}\eta^\a_a.$$

We also write the superconformal transformations of the harmonics
\bea
&&\de_{sc}U^{+k}_a=\la^{+k}U^0_a-\la^{++}U^{-k}_a,\q \de_{sc}U^0_a=
-\la^{+k}U^-_{ka},\q\de_{sc}U^{-k}_a=0,\nn\\
&&\la^{+k}=ik_\ab\te^{+k\a}\te^{0\bt}+U^{+k}_aU^0_b\om_{ab}
+i(\te^{+k\a}U^0_a-\te^{0\a}U^{+k}_a)\eta_{a\a},\lb{la1}\\
&&\la^{++}=\sfrac{i}2k_\ab(\te^{+l\a}\te^{+\bt}_l)+\sfrac12U^{+k}_a
U^+_{kb}\,\om_{ab}+i\te^{+k\a}U^+_{ka}\eta_{a\a}=\sfrac12\cD^+_k\la^{+k},
\nn\\
&&\cD^{+i}\la^{+k}=\ve^{ik}\la^{++},\q \cD^{+i}\la^{++}=0,\q\cD^i_l\la^{+k}
=\de^k_l\la^{+i},\q\cD^i_l\la^{++}=\de^i_l\la^{++}.
\eea

It is easy to obtain expressions for the $SC^5_3$ generators in
the analytic superspace; for instance, the generator of the special conformal
transformations has the form
\bea
&&K_m^\A=y_my^n\pa_n-\sfrac12y^2\pa_m+\sfrac12y^n\te^{0\bt}
(\g_m)_{\bt\rh}(\g_n)^{\rh\a}\pa^0_\a\nn
\\
&&+\sfrac12y^n\te^{+k\bt}(\g_m)_{\bt\rh}(\g_n)^{\rh\a}\pa^-_{k\a}
+\sfrac{i}4(\te^0)^2\te^{+k\bt}(\g_m)_\bt^\a\pa^-_{k\a}\nn\\
&&-i(\te^{+k}\g_m\te^0)\pa^-_k+\sfrac{i}2(\te^{+l}\g_m\te^+_l)\pa^{--}.
\eea
\\

{\bf A.3. $SC^5_3$-transformations of analytic measure}\\

Analytic integration measure \p{meanal} is invariant with respect to
the $SC^5_3$ transformations
\bea
&&\de_{sc}d\mu^{-4}=d\mu^{-4}(\pa_m\de_{sc}y^m+\frac{\pa}{\pa U^{0}_a}
\de_{sc}U^0_a+\frac{\pa}{\pa U^{+k}_a}\de_{sc}U^{+k}_a\nn\\&&
-\pa^0_\a\de_{sc}\te^{0\a}-\pa^-_{k\a}\de_{sc}\te^{+k\a})=0.\lb{scmu}
\eea
Here we use the formulas
\bea
&&\pa_m\de_{sc}y^m=3k_my^m+6l+\sfrac{3i}2\te^{0\a}\eta_\a^{0}
+3i\te^{+k\a}\eta_{k\a}^{-},\nn\\
&&\frac{\pa}{\pa U^{0}_a}
\de_{sc}U^0_a+\frac{\pa}{\pa U^{+k}_a}\de_{sc}U^{+k}_a=
-2\om_{ab}U^{+k}_aU^-_{kb}-2i\te^{0\a}\eta^0_{\a}
-2i\te^{+k\a}\eta^-_{k\a},\\
&&-\pa^0_\a\de_{sc}\te^{0\a}-\pa^-_{k\a}\de_{sc}\te^{+k\a}=
-6l-3y^mk_m+2\om_{ab}U_a^{+k}U_{kb}^{-}+\sfrac{i}2\te^{0\a}\eta^{0}_\a
-i\te^{+k\a}\eta^{-}_{k\a},\nn
\eea
derived by  differentiating the infinitesimal transformations
of the analytic coordinates.
\\

{\bf A.4. $SC^5_3$-transformations of harmonic derivatives}\\

The $SC^5_3$-transformations of the analytic coordinate generate
the superconformal transformations of the harmonic derivatives \p{ABharm}
\bea
&&\de_{sc}\cD^{+k}=-\la^{+l}\cD^k_l,\q\de_{sc}\cD^{++}
=\la^{++}\cD^k_k-\la^{+}_l\cD^{+l},\lb{scharm}\\
&&\de_{sc}\cD^-_k=(\cD^-_k\la^{+l})\cD^-_{l}+(\cD^-_k\la^{++}
-\la^{+}_k)\cD^{--},\nn\\
&&\de_{sc}\cD^{--}=(\cD^{--}\la^{+l})\cD^-_{l}+(\cD^{--}\la^{++})
\cD^{--},\q\de_{sc}\cD^k_l=0,
\eea
where the superfield parameters  $\la^{+l}, \la^{++}$ are defined in
the transformations of harmonics \p{la1}.
\\

{\bf A.5. Algebra of spinor coordinates}\\

We consider the  independent combinations of the Grassmann
coordinates $\te^{+k\a}$
\bea
&&(\te^{+k}\te^{+l})=\te^{+k\a}\te^{+l}_\a=(\te^{+l}\te^{+k}),\q
(\te^{+\a}\te^{+\bt})=\te^{+k\a}\te^{+\bt}_k=(\te^{+\bt}\te^{+\a}),\nn\\
&&\T^{+3l\a}\equiv\sfrac13(\te^{+k}\te^{+l})\te^{+\a}_k=-\sfrac13\te^{+l}_\bt(\te^{+\a}
\te^{+\bt}),\\
&&\T^{(+4)}=\sfrac13(\te^{+k}\te^{+l})(\te^{+}_k\te^{+}_l)
=(\te^{+1})^2(\te^{+2})^2
=-\sfrac13(\te^{+\a}\te^{+\bt})(\te^{+}_\a\te^{+}_\bt).\nn
\eea

Including the neutral spinor coordinate, we can also obtain  the
 combinations
\bea
&&(\te^0)^2=\te^{0\a}\te^0_\a,\q (\te^{+i}\te^0)=\te^{+i\a}\te^0_\a,\q
(\te^{+i}\g^m\te^0)=-(\te^0\g^m\te^{+i}),\nn\\
&&\T^{+i\a}=(\te^0)^2\te^{+i\a},\q \T^{++ik\a}=(\te^{+i}\te^{+k})\te^{0\a}
,\\
&&\T^{++\a}=\te^{+k\a}\te^{+\bt}_k\te^{0}_\bt,\q
\T^{++(\a\bt\g)}=(\te^{+(\a}\te^{+\bt})\te^{0\g)},\nn
\eea
where the last quantity is totally symmetric in spinor indices. The
special conjugation of the Grassmann polynomials is defined by
 \p{spconj}
\bea
&&(\te^{+k}\te^{+k})^\sim=-(\te^{+k}\te^{+l}),\q
(\te^{+\a}\te^{+\bt})^\sim=-(\te^{+\a}\te^{+\bt}),\q
[(\te^0)^2]^\sim=-(\te^0)^2,\nn\\
&&(\te^{+i}\te^0)^\sim=-(\te^{+i}\te^0),\q(\te^{+i}\g^m\te^0)^\sim
=-(\te^{+i}\g^m\te^0),\q(\T^{(+4)})^\sim=\T^{(+4)},\\
&&(\T^{+i\a})^\sim=-\T^{+i\a},\q (\T^{++\a})^\sim=-\T^{++\a},\q
(\T^{++(\a\bt\g)})^\sim=\T^{++(\a\bt\g)}.\nn
\eea

\end{document}